\documentstyle[preprint,aps]{revtex}
\begin{document}
\title{Do the $A_4$C$_{60}$ fullerides have a broken-symmetry ground state?}
\author{S.C. Erwin}
\address{Department of Physics, University of Pennsylvania, Philadelphia,
Pennsylvania 19104}
\author{C. Bruder}
\address{Theoretische Festk\"{o}rperphysik, University of Karlsruhe,
76128 Karlsruhe, Germany}
\date{August 11, 1993}
\maketitle
\begin{abstract}
Band theory predicts both K$_3$C$_{60}$ and K$_4$C$_{60}$ to be
metals; various experimental probes show that while K$_3$C$_{60}$ is
indeed metallic, K$_4$C$_{60}$ appears to be insulating.  The standard
view of this apparent failure of the single-particle picture is that
electron correlation is predominant.  We describe an alternative
scenario, motivated on theoretical grounds, which invokes a spin- or
charge-density-wave state to explain the observed insulating behavior.
\vspace*{0.25in}
\begin{flushleft}
(To appear in Physica B)
\end{flushleft}

\end{abstract}


\newpage
\section{Introduction}

Since the discovery of superconductivity in the potassium-intercalated
fulleride K$_3$C$_{60}$, considerable theoretical attention has
focused on the electronic properties of the $A_n$C$_{60}$ family
($A$=alkali), much of it assuming the validity of a single-particle
band description. A growing body of evidence suggests that this
assumption may rest on rather shaky ground, at least for some members
of the family.  A striking example is the contrast between
K$_3$C$_{60}$ and K$_4$C$_{60}$.  According to first-principles band
calculations, both are metallic, with comparable Fermi-level densities
of states.  Photoemission gives a very different picture, however: for
K$_3$C$_{60}$, strong Fermi-level emission is observed, while for
K$_4$C$_{60}$ the spectrum is insulating \cite{weaver}.  Additional
evidence is found in NMR, which also shows that K$_3$C$_{60}$ is
metallic and K$_4$C$_{60}$ is insulating \cite{murphy}.  Several
workers have argued that strong electron-electron correlation splits
the conduction band (CB) into a filled lower Hubbard band and an empty
upper Hubbard band \cite{benning}.

In this paper, we propose a different scenario to explain the
insulating nature of the $A_4$C$_{60}$ fullerides.  We begin by
discussing the results of first-principles band-structure calculations
for K$_4$C$_{60}$, then explore possible broken-symmetry ground states
such as a charge-density wave (CDW).  We emphasize that our results do
not in any sense {\it prove} that a CDW state is the thermodynamic
ground state; rather, we make a plausibility argument and then
describe model-Hamiltonian calculations which may give further
insight.

\section{Band structure of K$_4$C$_{60}$}

We use the local-density approximation (LDA) to density-functional
theory.  Bloch basis functions are linear combinations of occupied and
unoccupied atomic orbitals for potassium and carbon; these are in turn
expanded on a set of gaussian functions. This provides a compact
basis, and allows us to perform accurate all-electron
calculations---without the need for pseudopotentials---in which core,
valence, and conduction states are treated on equal footing. The
potential and charge density are completely general and without shape
approximation. We have previously used identical methods to study
K$_6$C$_{60}$ \cite{erwin1}, K$_3$C$_{60}$ \cite{erwin2}, and
Ba$_6$C$_{60}$ \cite{erwin3}; further details concerning numerical
methods may be found in these references, and preliminary work on
$A_4$C$_{60}$ may be found in Ref.~\cite{erwin4}.

The K$_4$C$_{60}$ geometry used in our calculation is taken from the
x-ray refinement of Fleming {\it et al.} \cite{fleming}, who found a
body-centered tetragonal (bct) lattice for the $A_4$C$_{60}$ family,
with lattice constant $a$=11.886 \AA~and $c/a$=0.906 for
$A$=potassium.  The potassium positions were refined to
(0.22,0.50,0)$a$ and symmetry-related sites.  Fleming {\it et al.}
observed that because C$_{60}$ lacks a four-fold symmetry axis, a
tetragonal lattice ($a=b$) implies that the C$_{60}$ molecules are
either orientationally disordered or are spinning rapidly; Zhou and
Cox have shown that a merohedrally disordered model fits the data well
\cite{zhou}.  Rabe has proposed an alternative structural model
\cite{rabe}, consisting of an orientationally ordered crystal with a
screw-axis symmetry that leads naturally to $a=b$. Here, we make the
standard simplification of taking the K$_4$C$_{60}$ structure to be
orientationally ordered, with one C$_{60}$ molecule per cell.  Work
currently in progress, which goes beyond this simplification, will be
described in Sec.~\ref{hubbard}.

In Fig.~\ref{bands} we show the band structure for orientationally
ordered K$_4$C$_{60}$. Since the bct conventional cell is a slightly
distorted version of the body-centered-cubic (bcc) cell of
K$_6$C$_{60}$, we show the bands along the bct route that corresponds
most closely to the bcc route appearing in Fig.~2 of
Ref.~\cite{erwin1}.  Indeed, the K$_4$C$_{60}$ bands bear a close
resemblance to those of K$_6$C$_{60}$, with an important difference.
The point group of orientationally ordered $A_4$C$_{60}$ (generated
from three Cartesian mirror planes) has only one-dimensional
irreducible representations, so that any band degeneracies must be
accidental.  This has the effect of smearing somewhat the spectral
density throughout the CB, resulting in a DOS with few sharp features.
The Fermi-level DOS $N(\epsilon_F)$=18.3 states/eV-cell (both spins)
is $\sim$40\% larger than for K$_3$C$_{60}$, strongly at variance with
the NMR and photoemission results discussed above.

The Fermi surface (FS) has 3 sheets, shown in Fig.~\ref{fs}.  Two form
closed hole pockets centered at $\Gamma$.  The third forms
quasi-two-dimensional sheets, large parts of which are nearly flat
over much of the zone and roughly normal to the $c$-axis.  To the
extent that these sheets are truly flat, they represent an unusual
form of FS nesting, with a nesting vector ${\bf Q}=Q\hat{z}$.  It is
well known that FS nesting can lead to electronic instabilities, by
enhancing the susceptibility $\chi({\bf Q})$ for the formation of a
charge- or spin-density-wave (SDW) ground state. The simplest
illustration of this occurs in a linear chain of atoms equally spaced
by $a$, for which a half-filled cosine band has a nesting vector equal
to half the zone, $Q=\pi/a$. The Peierls mechanism provides a route to
lowering the total energy, by chain dimerization (that
is, by forming a commensurate CDW with period $2a$) and the opening of
a gap.

To quantify the degree of nesting and identify posible nesting vectors
in K$_4$C$_{60}$, we calculate
\begin{equation}
\xi_{nm}({\bf Q}) = N^{-1} \sum_{\bf k} \delta(\epsilon_{n,{\bf
k}}-\epsilon_F)
 \delta(\epsilon_{m,{\bf k}+{\bf Q}}-\epsilon_F),
\end{equation}
which measures the phase space available for any process that scatters
electrons from point ${\bf k}$ on the $n$th FS sheet to ${\bf k}+{\bf
Q}$ on the $m$th sheet. This factor also enters into the second-order
susceptibility, $\partial^2 E_t/\partial h_{\bf Q}
\partial h_{-{\bf Q}}$, of the total energy $E_t$ with respect to
density variations $h_{\bf Q}$, making explicit the connection between
nesting vectors and CDW wave vectors.  In Fig.~\ref{xi} we plot the
function $\xi_{33}({\bf Q})$ in the $Q_x,Q_z$ plane, revealing a
single moderately strong nesting vector centered near ${\bf
Q}=\pm(0,0,\pi/c)$.  This is precisely the nesting vector required for
the smallest possible commensurate density wave, namely, period
doubling along the $c$-axis.  In the following section we describe
preliminary work, using a model Hamiltonian, that may lend further
insight into this scenario.

\section{Mean-field Hubbard model}\label{hubbard}

Shortly after the discovery of superconducting K$_3$C$_{60}$, Zhang,
Ogata, and Rice proposed a negative-$U$ model for fulleride
superconductivity mediated by alkali-ion optical phonons \cite{zhang}.
They argued that {\it if} the interaction is attractive, then
superconductivity would result from the competition between a CDW
ground state and a superconducting ground state, for which the fcc
lattice of K$_3$C$_{60}$ frustrates the former in favor of the latter.
Since this proposal, isotope experiments have demonstrated that alkali
optical phonons are not involved in pairing.  Moreover, considerable
experimental and theoretical evidence has shown that the
electron-electron interaction is intrinsically {\it repulsive}.
Nevertheless, we discuss the negative-$U$ model here for two reasons:
(1) we will later use the same formalism to study the consequences of
a repulsive interaction, and (2) it provides a pedagogically
instructive context for a theorem which states that for the $U<0$
Hubbard model with a single band and nearest-neighbor hopping, the CDW
and superconducting ground states are exactly degenerate when the band
is half filled {\it and} the lattice is bipartite
\cite{nagoaka}.  For K$_3$C$_{60}$, the band is half filled but the
lattice is not bipartite; the authors of Ref.~\cite{zhang} showed
numerically that the degeneracy is broken in favor of a
superconducting ground state.  For K$_4$C$_{60}$, the situation is
reversed: the lattice is bipartite but the filling is two-thirds,
again breaking the degeneracy. It is tempting to hypothesize that
since (1) the bct lattice is bipartite, and hence no longer obviously
frustrates the CDW state, and (2) the ratio $c/a<1$ provides a
preferred direction for the CDW wave vector, then perhaps the
degeneracy might be broken in favor of the CDW state.

To explore this possibility, we proceed along similar lines to
Ref.~\cite{zhang}, solving the negative-$U$ Hubbard model within the
mean-field approximation.  We begin with the Hamiltonian
\begin{equation}
H = -t \sum_{\left\langle i,j \right\rangle} \left( c^\dagger_i c_j +
h.c. \right) + U \sum_i n_{i\uparrow} n_{i\downarrow}.
\label{hhubbard}
\end{equation}
We assume a single band with nearest-neighbor hopping on the bct
lattice; the dispersion is then
\begin{equation}
\beta_{\bf k} = -8\cos(k_xa/2)\cos(k_ya/2)\cos(k_zc/2).
\end{equation}
For the CDW state, the energy per electron is
\begin{equation}
E^{CDW}=
\frac{1}{\zeta N} \sum_{{\bf k},s=\pm1}
\epsilon^{CDW}_{{\bf k},s}  +
\frac{\zeta U}{4} \left[1-(\delta/\zeta)^2\right],
\label{ecdw}
\end{equation}
where $\delta$ is the order parameter given by $\langle n_i
\rangle=1\pm\delta$, and $\zeta=4/3$ is (twice) the band filling. The
sum is over filled states and the band energy is given by
\begin{equation} \epsilon^{CDW}_{{\bf k},s}=
s\left[ (t\beta_{\bf k})^2 + (U\delta/2)^2 \right]^{1/2}.
\end{equation}
For the superconducting
state (SS), the energy per electron is
\begin{equation}
E^{SS}=
-\frac{1}{\zeta N} \sum_{\bf k}
\epsilon^{SS}_{\bf k} + \epsilon_F +
\frac{\zeta U}{4} \left[1-(2\Delta/\zeta)^2 \right],
\label{ess}
\end{equation}
where $\Delta=\langle c^\dagger_{i\uparrow}
c^\dagger_{i\downarrow} \rangle$
is the superconducting order parameter, and the band energy is now
\begin{equation}
\epsilon^{SS}_{\bf k}=
\left[ (\beta_{\bf k}-\epsilon_F)^2 + (U\Delta)^2 \right]^{1/2}.
\end{equation}
In both Eqs.~(\ref{ecdw}) and (\ref{ess}), the total energy is
evaluated by minimization with respect to the order parameter.

We have calculated the CDW and SS energies for $0\leq|U|/t\leq10$; the
model predicts the SS state to be favored throughout this range.  This
is strongly at odds with experiment (which finds no evidence for
superconductivity) and with our hypothesis of a favored CDW state. We
discuss this below.

Although instructive, the Hamiltonian of Eq.~(\ref{hhubbard}) is too
crude in several important ways: (1) A single tight-binding band does
not reproduce the nesting features of Fig.~\ref{fs}. (2) Recent work
on tight-binding parameterizations of the LDA energies shows that
nearest-neighbor interactions alone are inadequate for reproducing the
LDA spectrum, and that two coordination shells are necessary
\cite{krishna}. (3) A variety of experimental and theoretical evidence
has led to a consensus that $U$ is {\it positive}, with a value of
order 1 eV.  To address these deficiencies, we have begun to study the
Hamiltonian
\begin{equation}
H = \sum_{i\mu,j\nu} \left( c^\dagger_{i\mu}
T_{i\mu,j\nu}
c_{j\nu} +
h.c. \right) + U \sum_i n_{i\uparrow} n_{i\downarrow},
\end{equation}
where $c^\dagger_{i\mu}$ creates an electron on site $i$ in orbital
$\mu=x,y,z$.  The parameter $t$ has been generalized here to a set of
3$\times$3 matrices giving the amplitude for hopping between different
orbitals on different sites (possibly between molecules in different
orientations).  We retain hopping terms between first and second
neighbors.  The matrix elements $T_{i\mu,j\nu}$ are obtained from the
LDA eigenvalue spectrum by Fourier inversion, giving the optimal
tight-binding representation of the LDA dispersion.  Finally, we study
both positive- and negative-$U$ solutions, i.e., both SDW and CDW
states. Preliminary indications are of an enhanced susceptibility for
forming a density-wave state with ${\bf Q}$ oriented along the
$c$-axis, corresponding to a value of $|U|\sim 1$ eV \cite{krishna}.

\section{Period doubling and band structure}

A complementary line of inquiry can be followed within the
single-particle picture. LDA has previously been used to study
commensurate CDW's by first assuming, say, a doubled unit cell and
then minimizing the total energy with respect to a dimerization
coordinate.  Whether or not this procedure works in principle is the
subject of debate: Overhauser has argued that correlation effects are
important for density-wave ground states and are too crudely represented
by LDA \cite{overhauser}, while Ashkenazi {\it et al.} have shown
numerically that LDA fails to predict {\it any} dimerization even for
the textbook Peierls system, polyacetylene \cite{ashkenazi}.

For these reasons, we focus on band-structure effects.  We assume that
a density wave in K$_4$C$_{60}$ leads to period-doubling along the
$c$-axis (as suggested by the FS nesting), with an amplitude of 5\%.
The orientations of the C$_{60}$ molecules are assumed not to change.
The resulting LDA band structure for this doubled cell is suggestive,
but not definitive: no gap appears, but the extended flat FS sheet
disappears, leaving only small electron and hole pockets.  We
estimate, from simple FS-area scaling arguments, that the dc
conductivity is smaller by a factor of 5--10.  While not insulating
{\it per se}, the reduced conductivity may be manifested in NMR as a
small Korringa component plus an activated component.  Additional
effects from disorder (orientational or defects) are expected to reduce
further the measured conductivity.\\

We acknowledge helpful discussions with E.J. Mele and W.E. Pickett.

\begin{figure}
\caption{Theoretical LDA band structure for (orientationally ordered)
K$_4$C$_{60}$.  The Fermi level is the energy zero.
\label{bands}}
\end{figure}

\begin{figure}
\caption{Fermi surface corresponding to the LDA bands of Fig.~1,
in a repeated-zone representation.
\label{fs}}
\end{figure}

\begin{figure}
\caption{Surface plot of $\xi_{33}({\bf Q})$ in the $Q_x,Q_z$ plane.
The plotting rectangle is centered on ${\bf Q}=0$, and has sides of length
$3(\pi/a)$ and  $3(\pi/c)$ in the $Q_x$ and $Q_z$ directions,
respectively. The singularity at ${\bf Q}=0$ is physically meaningless.
\label{xi}}
\end{figure}


\begin{references}

\bibitem {weaver} J.H. Weaver and D.M. Poirier, in {\it Solid State Physics},
Volume {\bf 48}, H. Ehrenreich and F. Spaepen, eds. (Academic, New York, 1994).
\bibitem {murphy} D.W. Murphy {\it et al.}, preprint.
\bibitem {benning} P.J. Benning {\it et al.}, Phys.~Rev.~B~{\bf 47}, 13843
(1993).
\bibitem {erwin1} S.C. Erwin and M.R. Pederson, Phys.~Rev.~Lett.~{\bf 67}, 1610
(1991).
\bibitem {erwin2} S.C. Erwin and W.E. Pickett, Science {\bf 254}, 842 (1991).
\bibitem {erwin3} S.C. Erwin and M.R. Pederson, Phys.~Rev.~B~{\bf 47}, 14657
(1993).
\bibitem {erwin4} S.C. Erwin, in {\it Buckminsterfullerenes}, W.E. Billups and
M.A. Ciufolini, eds. (VCH, New York, 1993) p.~217.
\bibitem {fleming} R.M. Fleming {\it et al.}, Nature {\bf 352}, 701(1991);
errata.
\bibitem {zhou} O. Zhou and D. Cox, J. Phys.~Chem.~Solids {\bf 53}, 1373
(1992).
\bibitem {rabe} K.M. Rabe, preprint.
\bibitem {zhang} F.C. Zhang, M. Ogata, and T.M. Rice, Phys. Rev. Lett. {\bf
67}, 3452 (1991).
\bibitem {nagoaka} Y. Nagaoka, Prog.~Theor.~Phys.~{\bf 52}, 1716 (1974).
\bibitem {krishna} V. Krishna, S.C. Erwin, E.J. Mele, and C. Bruder (to be
published).
\bibitem {overhauser} A.W. Overhauser, in {\it Charge Density Phenomena in
Potassium}, J.T. Devreese and F. Brosens, eds. (Plenum, New York, 1983), p.~41.
\bibitem {ashkenazi} J. Ashkenazi {\it et al.}, Phys.~Rev.~Lett.~{\bf 62}, 2016
(1989).

\end{references}
\end{document}